\documentclass[%
 reprint,
superscriptaddress,
 amsmath,amssymb,
 aps,prl
]{revtex4-2}

\usepackage{dsfont}
\usepackage{graphicx}
\usepackage{dcolumn}
\usepackage{bm}
\usepackage{hyperref}
\hypersetup{
colorlinks=true,
	linkcolor=nrppurple,
	citecolor=nrppurple,
	urlcolor=nrppurple,
}

\usepackage{ulem} 

\newcommand{\del}{\partial}
\renewcommand{\v}[1]{\bm{\mathrm{#1}}}
\newcommand{\uv}[1]{\v{\hat{#1}}}
\renewcommand{\k}{\v{k}}
\newcommand{\q}{\v{q}}

\newcommand{\epsi}{\varepsilon}

\newcommand{\tr}{\mathrm{tr}}
\newcommand{\id}{\mathds{1}}

\newcommand{\pfrac}[2]{\left( \frac{#1}{#2} \right)}

\newcommand{\goes}{\rightarrow}

\newcommand{\comment}[1]{} 

\newcommand{\e}{\mathrm{e}}
\newcommand{\w}{\omega}
\newcommand{\W}{\Omega}

\makeatletter \renewcommand\@make@capt@title[2]{%
\@ifx@empty\float@link{\@firstofone}{\expandafter\href\expandafter{\float@link}}%
\sffamily{\textbf{#1}}\@caption@fignum@sep#2 }
\usepackage{placeins} 

\begin{document}
\definecolor{nrppurple}{RGB}{128,0,128}

\preprint{APS/123-QED}

\title{Surface Cooper pair spin waves in triplet superconductors}

\author{Nicholas R. Poniatowski}
\email[]{nponiatowski@g.harvard.edu}
\affiliation{Department of Physics, Harvard University, Cambridge, MA 02138, USA}
\author{Jonathan B. Curtis}
\affiliation{Department of Physics, Harvard University, Cambridge, MA 02138, USA}
\affiliation{John A. Paulson School of Applied Sciences and Engineering, Harvard University, Cambridge Massachusetts 02138 USA}
\author{Charlotte G.L. B\o ttcher}
\affiliation{Department of Physics, Harvard University, Cambridge, MA 02138, USA}
\author{Victor M. Galitski}
\affiliation{Joint Quantum Institute, Department of Physics, University
of Maryland, College Park, MD 20742, USA}
\author{Amir Yacoby}
\affiliation{Department of Physics, Harvard University, Cambridge, MA 02138, USA}
\affiliation{John A. Paulson School of Applied Sciences and Engineering, Harvard University, Cambridge Massachusetts 02138 USA}
\author{Prineha Narang}
\affiliation{John A. Paulson School of Applied Sciences and Engineering, Harvard University, Cambridge Massachusetts 02138 USA}
\author{Eugene Demler}
\affiliation{Institute for Theoretical Physics, ETH Z\"urich, 8093 Z\"urich, Switzerland}

\date{\today}

\begin{abstract}
    We study the electrodynamics of spin triplet superconductors including dipolar interactions, which give rise to an interplay between the collective spin dynamics of the condensate and orbital Meissner screening currents. Within this theory, we identify a class of spin waves that originate from the coupled dynamics of the spin-symmetry breaking triplet order parameter and the electromagnetic field. In particular, we study magnetostatic spin wave modes that are localized to the sample surface. We show that these surface modes can be excited and detected using experimental techniques such as microwave spin wave resonance spectroscopy or nitrogen-vacancy magnetometry, and propose that the detection of these modes offers a means for the identification of spin triplet superconductivity. 
\end{abstract}

\maketitle

Spin triplet superconductors are distinguished from their conventional spin singlet counterparts by the fact that they spontaneously break spin-rotation symmetry in addition to global $U(1)$ symmetry. This additional symmetry breaking implies the existence of collective modes, analogous to spin waves in other spin-symmetry breaking systems such as ferromagnets and antiferromagnets \cite{basicnotions}, which originate from the coherent precession of the triplet order parameter. Therefore, the existence of these modes, manifested in the spin dynamics of the ordered state, can be used to experimentally identify spin triplet superconductors. In fact, this approach proved to be spectacularly successful in the study of superfluid $^3$He, where probes of the spin dynamics, such as NMR, were crucial in the identification of the triplet order parameter \cite{3he-book,volovik-book,leggett-rmp}. Along these lines, there have been several proposals to probe spin waves in triplet superconductors via thermodynamic \cite{eugene-spinwaves,214-modes,maki-214} and transport \cite{214-spincurrent} measurements, but to date these modes have not been observed in any solid state system. Further, there has been extensive work concerning orbital collective modes in unconventional superconductors \cite{power-1,power-2,Yip1992,214-1d,valentin-1,214-anisotropic,chiral-higgs,clappingmodes} and their (as yet unrealized) potential use as hallmarks of multi-component condensates. 

In contrast, spectroscopic probes which rely on the coupling of spin waves to electromagnetic fields have proven to be a powerful tool in the study of  magnetically ordered systems \cite{kittel,FMR-review}. At the long wavelengths relevant to experiments, spin dynamics are dominated by dipolar interactions, which can lead to the emergence of collective modes corresponding to the coupled fluctuations of spin waves in the material and incident electromagnetic fields \cite{sw-book,walker,DE-original}. The most notable example is the Damon-Eshbach mode in ferromagnets \cite{DE-original} and antiferromagnets \cite{afm-de-1,afm-de-2}, which is localized to the sample surface. 

In this work, we develop an electrodynamic theory of spin waves in triplet superconductors, including the effects of dipolar interactions which couple the spin and orbital dynamics of the condensate. Physically, the precession of the triplet order parameter generates a magnetic field that couples back onto the condensate magnetization, supporting collective dipolar modes. These modes remain stable even when coupled to the orbital motion of the condensate through Meissner screening. At the sample surface, this effect gives rise to a distinct set of surface modes, analagous to Damon-Eshbach modes, which we show are especially promising for experimental detection.
The collective modes are identified from the solutions of Maxwell’s equations in the triplet medium, characterized by a dynamical magnetic susceptibility $\underline{\chi}(\W)$. Thus, we must begin by developing the theory of the uncoupled spin dynamics of the triplet system. 

\textit{Matter dynamics  -- } A spin triplet superconductor is characterized by an $S = 1$ order parameter which is a symmetric matrix in spin space. It is conventional to represent this matrix in terms of the so-called ``\v{d}-vector'' as \cite{3he-book,leggett-rmp}
\begin{equation}
    \hat{\Delta} = \begin{pmatrix} \Delta_{\uparrow \uparrow} & \Delta_{\uparrow \downarrow} \\ \Delta_{\downarrow \uparrow} & \Delta_{\downarrow \downarrow} \end{pmatrix} = \begin{pmatrix} -d_x + i d_y & d_z \\ d_z & d_x + id_y \end{pmatrix} \, .
\end{equation}
Physically, one may envision $\uv{d}$ as the direction along which the condensate has an $m=0$ spin projection.
In rare cases, the condensate may be ``non-unitary,'' and possess a net spin polarization given by $\v{Q} = i\v{d}\times \bar{\v{d}}$ \cite{sigrist-rmp}.

To describe the low energy behavior of a triplet superconductor, we may construct a phenomenological Ginzburg-Landau free energy,
\begin{equation} \label{eff-th}
\begin{split}
   \mathcal{F} =  \int d^3 r \left[ - r \bar{\v{d}}\cdot \v{d} - \Gamma \bar{d}_z d_z 
    + \frac{u}{2} \big(\bar{\v{d}}\cdot \v{d} \big)^2 + \frac{\tilde{u}}{2} \v{Q}^2 - g\v{Q}\cdot \v{h}\right] \, , 
\end{split}
\end{equation}
where $r,u,\tilde{u},g$ are coupling constants and we have neglected gradient terms \footnote{Gradient terms are sub-dominant at long-wavelengths compared to the orbital Meissner screening currents. Including gradient terms, the magnetostatic mode dispersion acquires a positive group velocity at short wavelengths $\lambda q > 1$, in direct analogy to the inclusion of exchange terms in the magnetostatic theory of ferromagnets and antiferromagnets.}.
We have introduced a phenomenological easy-axis spin anisotropy, $\Gamma > 0$ that pins the $\v{d}$ vector along the $\uv{z}$ axis \footnote{In general, $\v{d}(\v{r},t)$ is a function of momentum. However, as a consequence of the $\v{d}$-vector pinning, we assume that $\v{d}(\v{r},t)$ is uniformly oriented along $\uv{z}$ at all points on the Fermi surface, and thus independent of the relative momentum of the triplet pairs.}. 
The sign of $\tilde{u}$ determines whether or not an equilibrium condensate magnetization is favored, and in what follows we will take $\tilde{u} > 0$ to study unitary  states, since spin-polarized non-unitary states are generally energetically unfavorable and thus exceedingly rare. Finally, we have coupled the condensate magnetic moment $g \mathbf{Q}$ to an external magnetic field, $\v{h}$. Although $\v{Q}$ vanishes in equilibrium for unitary states, the energetics of sustaining a fluctuating $\v{Q}$ out-of-equilibrium affects the spin dynamics, as will be shown below. We consider only applying a weak field (which does not affect the equilibrium $\v{d}$-vector) as a means to probe the system. We note that this model is essentially equivalent to that of a spin-one spinor condensate \cite{spinor-bec}. 

Considering a static system and minimizing Eq. (\ref{eff-th}), the equilibrium $\v{d}$-vector is found to be $\v{d}_0 = v \, \uv{z}$, with $v = \sqrt{(r+\Gamma)/u}$. To study the order parameter fluctuations, we parameterize the six real collective modes of the triplet superconductor as
\begin{equation}
    \v{d} = \e^{i\varphi} \, \big[\big( v+ h) \uv{z} + \v{a} + i \v{b} \big] \, ,
\end{equation}
where the global phase mode $\varphi$ is lifted to the plasma frequency by the Anderson-Higgs mechanism \cite{anderson-higgs}, and the longitudinal fluctuation of the $\v{d}$-vector is the usual amplitude Higgs mode which resides at the gap edge \cite{higgs-arcmp}. Additionally, there are four transverse modes $\v{a} = (a_x,a_y,0)$ and $\v{b} = (b_x,b_y,0)$ which are respectively in- and out-of-phase with the equilibrium condensate. Physically, these modes are two species of spin waves associated with the coherent fluctuation of the superconducting condensate, i.e. Cooper pair spin waves. 

By applying the $S=1$ representations of the generators of $SU(2)$ spin rotations to $\v{d}_0$, we see that the $\v{a}$ mode corresponds to long-wavelength rotations of the $\v{d}$-vector. In the absence of anisotropy, $\v{a}$ is a Goldstone mode, while a finite anisotropy $\Gamma$ gaps the mode to $\W_{a} = \Gamma$. 
In contrast, the $\v{b}$ mode gives rise to a fluctuating magnetization $\delta \v{Q} = 2 \mathbf{d}_0 \times \mathbf{b}$, and is gapped to a frequency $\W_b = 2\tilde{u}v^2 + \Gamma$ since magnetization fluctuations around the unitary ground state are energetically costly. The $\v{a}$ and $\v{b}$ modes are analogous to the fluctuations of the Ne\'el vector and magnetization in an antiferromagnet, which can be thought of as in- and out-of-phase fluctuations of the sublattice magnetizations, respectively. 

\begin{figure}
    \centering
    \includegraphics[width=85mm]{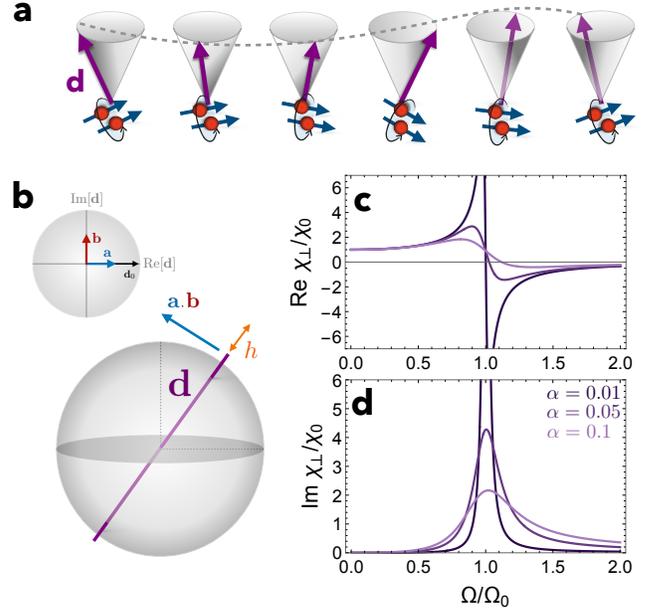}
    \caption{\textbf{Cooper pair spin waves.} \textbf{a.} Illustration of Cooper pair spin waves, represented as both the precession of the $\v{d}$-vector, and (schematically) of the associated triplet Cooper pair spins which arise due to the time-dependent transverse fluctuations of the $\v{d}$-vector. \textbf{b.} Illustration of the collective modes of a triplet superconductor: the longitudinal Higgs mode, $h$, and the transverse spin waves $\v{a}$ and $\v{b}$. The $\v{a}$ mode is in-phase with the equilibrium $\v{d}$-vector, while the $\v{b}$ mode is out of phase. \textbf{c.,d.} Dynamical transverse susceptibility $\chi_{\perp}(\W)$ for different values of the damping parameter $\alpha$ with $\chi_0 = 0.1$ and $\Gamma/\Omega_b = 0.05$. All curves exhibit resonant behavior at the Cooper pair spin wave frequency $\W_0^2 = \Gamma \W_b$.}
    \label{intro-fig}
\end{figure}

In order to study the collective dynamics of the triplet system, we employ a time-dependent Ginzburg-Landau formalism  ~\cite{dorsey-tdgl,h-h-rmp} which allows for both coherent and dissipative order parameter dynamics, with the equation of motion (taking $\hbar = 1$)
\begin{equation}
    i \del_t \v{d} = \frac{\delta \mathcal{F}}{\delta \bar{\v{d}}} + \alpha \del_t \v{d} \,
\end{equation}
where $\alpha$ is a dimensionless damping parameter that is analogous to the Gilbert damping parameter in the Landau-Lifshitz-Gilbert theory for magnetic dynamics. This phenomenological damping is meant to model the damping of spin waves by, e.g. nodal quasiparticles, which are otherwise absent in our purely bosonic theory. This Landau damping can be computed microscopically within weak coupling theory for a given model.

Solving the equations of motion, we can identify the magnetization $\v{m}= -\delta \mathcal{F}/\delta \v{h}$ and correspondingly the dynamic magnetic susceptibility $\v{m} = \underline{\chi} \v{h}$. 
Doing so, we find the transverse susceptibility $\chi_{xx} = \chi_{yy} = \chi_\perp$,
\begin{equation} \label{eff-susc}
    \chi_\perp(\W) = \chi_0 \, \frac{\W_b (\Gamma - i\alpha \W)}{(\Gamma - i\alpha \W)(\W_b - i\alpha \W) - \W^2 } \, .
\end{equation}
 The susceptibility takes the simple form $\chi_\perp(\W) = \chi_0 \W_0^2/(\W_0^2 - \W^2)$ in the absence of damping (with the resonant frequency $\W_0^2 = \Gamma \W_b$ and oscillator strength $\chi_0 = 2g^2v^2/\W_b$), and is plotted in Fig. \ref{intro-fig}{\color{nrppurple}c,d}. Like conventional magnetic systems, we take $\chi_0$ to coincide with the normal state paramagnetic susceptibility.
 
 This phenomenological model is meant to capture the key features of the low-frequency dynamics of a triplet condensate, in particular the pinning of the $\v{d}$ vector along an easy axis by spin-orbit coupling. 
Given a specific material, one can construct a more detailed model taking into account the relevant point group symmetry, but we expect the main results of our analysis to hold for purely triplet superconductors on the general grounds that the transverse susceptibility exhibits a resonance (associated with a bare spin wave excitation) at finite frequency \footnote{We also expect the magnetostatic modes studied in this work to be robust against order parameter domains, since these modes are achiral and determined by the transverse susceptibility. In practice, one expects that order parameter domains can be aligned with the application of a small ``training'' magnetic field, as demonstrated in Refs. \onlinecite{aharon-214,aharon-upt3,ian-ute2}}. The extension of our results to non-unitary or mixed-parity states is left as the subject of future work. 

 This generic form of the susceptibility can be illustrated by a  microscopic weak coupling calculation for a simple model system, which we describe in the Supplemental Information \footnote{\label{sicite} See Supplemental Information [url], which includes Refs. \onlinecite{odd-w,tsuneto,ll-v6,mw-sc-book,Gannon_2015,upt3-optics1} }. We consider a two-dimensional $p$-wave superconductor with an easy-axis spin anisotropy. At zero temperature, we find that the $\v{a}$ mode disperses according to $\W^2 = \W_0^2 + v_F^2 \v{q}^2/2$, where $v_F$ is the Fermi velocity and the Cooper pair spin wave frequency $\W_0^2 = \gamma_{\text{BCS}}(2\Delta_0)^2$ is set by a dimensionless measure of the spin anisotropy $\gamma_{\text{BCS}}$ (see the Supplemental Information) and the superconducting gap $2\Delta_0$. Meanwhile, the $\v{b}$ mode has the dispersion $\W^2 = (4\Delta_0)^2 + \W_0^2 + v_F^2 \v{q}^2/2$, which shows that $\W_b$ coincides with the gap edge in the spin-rotation symmetric system, and is pushed up into the quasiparticle continuum by anisotropy. Intrinsically, these modes disperse over electronic lengthscales, which can be safely neglected when studying the long-wavelength behavior of the system.  The transverse susceptibility \cite{Andreev_1980} in this model is $\chi_\perp(\W) = \chi_0 \, \W_0^2/(\W_0^2 - \W^2)$, consistent with the undamped susceptibility derived from time-dependent Ginzburg-Landau theory. This calculation also confirms that $\chi_0$ should be identified as the normal-state paramagnetic susceptibility.
 


\textit{Electrodynamics -- } Having derived the properties of the bare Cooper pair spin waves, we may now consider their interaction and hybridization with classical electromagnetic fields. Since the system is a superconductor, the diamagnetic response of the condensate, originating from orbital screening currents, must be taken into account when studying the spin wave electrodynamics. That is, the Cooper pair spin waves generate a fluctuating dipolar magnetic field, which must be screened by the Meissner currents. This unusual form of electrodynamics, where the spin dynamics and orbital currents are coupled at a classical level, distinguishes spin triplet superconductors from both ferromagnets and antiferromagnets as well as spinful, but uncharged, condensates such as superfluid $^3$He or atomic spinor Bose-Einstein condensates.

\begin{figure}
    \centering
    \includegraphics[width=75mm]{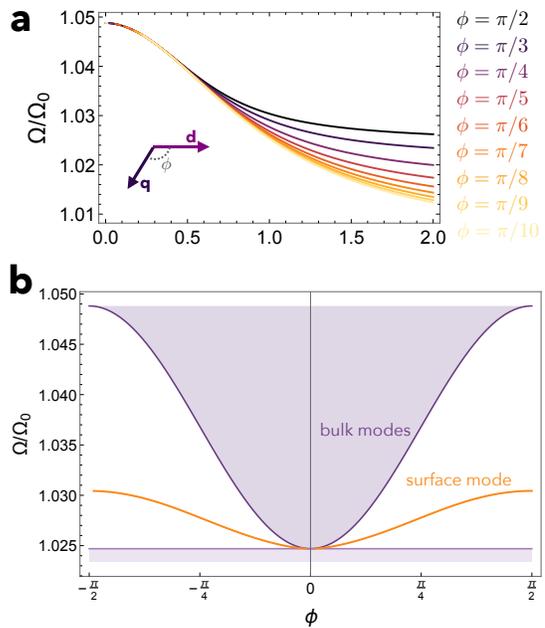}
    \caption{\textbf{Surface Cooper pair spin waves:} \textbf{a.} Dispersion of the surface mode for different directions of in-plane propagation with respect to the $\v{d}$-vector. \textbf{b.} Angular dependence of the bulk and surface mode frequencies for fixed in-plane wavevector $\lambda q = 1$. The surface-projected bulk spin wave bands form two continua, and the surface mode exists within the gap between them. }
    \label{de-fig}
\end{figure}

In the quasi-magnetostatic approximation, Maxwell's equations read (with $\mu_0 = 1$)
\begin{subequations} \label{magnetostatic-eq-1}
\begin{align}
    \nabla \cdot \v{b} &= 0 \, , \\
    \nabla \times \v{h} &= \v{j}_s \, ,
\end{align}
\end{subequations}
where $\v{j}_s$ is the screening supercurrent. This current must be conserved, and is governed by the London equation
\begin{subequations} \label{magnetostatic-eq-4}
\begin{align}
\nabla \cdot \v{j}_s &= 0 \, , \\
    \nabla \times \v{j}_s &= -\lambda^{-2} \, \v{b} \, , 
\end{align}
\end{subequations}
where $\lambda$ is the London penetration depth. Finally, in the triplet system, $\v{b}$ and $\v{h}$ are related by the constitutive relation $\v{b} = \underline{\mu} \v{h}$ with $\underline{\mu} = \underline{\id} + \underline{\chi}$ where the susceptibility (\ref{eff-susc}) encodes the bare spin wave spectrum. The magnetostatic collective modes are then the normal modes of the coupled Eq's. (\ref{magnetostatic-eq-1},\ref{magnetostatic-eq-4}), comprising of intertwined fluctuations of the magnetic field, Cooper pair spin waves, and induced supercurrent. 

\textit{Bulk modes --} We begin by considering an infinite triplet system, where we can solve the magnetostatic equations in momentum-space to find two branches of solutions. The first mode has the magnetic field polarized transverse to the $\v{d}$-vector and satisfies $\mu(\W) + \lambda^2 q^2 = 0$, where $\mu(\W) = 1 + \chi_\perp(\W)$ and $q$ is the wavevector of the mode. This mode is dispersive even in an infinite system on account of the Meissner effect. In addition to this transverse, isotropically dispersing mode, the second mode is anisotropic and partially longitudinal, satisfying $(1 + \lambda^2 q^2 \sin^2 \phi) \mu(\W) + \lambda^2 q^2 \cos^2 \phi = 0$, where $\phi$ is the angle between $\v{q}$ and the $\v{d}$-vector. The dispersions of the bulk modes are plotted in the Supplemental Information.

\textit{Surface modes --} Next, we consider a semi-infinite triplet superconductor occupying the half space $z <0$, with vacuum above ($z>0$). We take the $\v{d}$-vector to lie in the plane of the sample, along the $x$ axis. The magnetostatic equations (\ref{magnetostatic-eq-1},\ref{magnetostatic-eq-4}) must then be supplemented by the boundary conditions that $b_z, h_x$, and $h_y$ be continuous across the interface, and that $j_{s,z}$ must vanish at the interface to ensure supercurrent conservation. The details of this calculation are summarized in the Supplemental Information, and the key results are shown in Fig. \ref{de-fig}.

Figure \ref{de-fig}{\color{nrppurple}a} shows the dispersion of the surface mode for different directions of propagation with respect to the $\v{d}$-vector. Figure \ref{de-fig}{\color{nrppurple}b} shows the surface mode dispersion alongside the bulk spin wave bands projected onto the surface, which form two continua of scattering states as the out-of-plane wavevector $q_z$ is varied. As the direction of in-plane propagation becomes parallel to the $\v{d}$-vector ($\phi \goes 0$), the surface mode intersects the bulk spin wave bands and ceases to be a distinct, surface-localized mode. 

The Cooper pair spin wave frequency can be estimated for the candidate triplet superconductor UPt$_3$ \cite{upt3-rmp,upt3-sauls}, where NMR  \cite{upt3-pinning} and neutron scattering experiments \cite{gannon-upt3} show that the $\v{d}$-vector is depinned by a field of $\mu_0 H_{\text{pin}} = 230$ mT. We may then estimate the $\v{d}$-vector anisotropy to be $\Gamma \approx 2\mu_B \mu_0 H_{\text{pin}} \approx 300$ mK, and using the weak-coupling expression $2\Delta_0 \approx 3.5 k_B T_c \approx 1.75$ K, estimate the Cooper pair spin wave frequency to be $14$ GHz in this material. Broadly speaking, most candidate triplet superconductors have transition temperatures on the order of 1 K \cite{sheng-ute2,fm-sc,hillier-prl}. Estimating $\Gamma/2\Delta_0 \approx 5-20 \%$ \cite{sigrist-soc,soc-2}, we expect Cooper pair spin waves to generally reside at 15-30 GHz. However, these modes may be much lower-lying in organic superconductors \cite{eugene-spinwaves} or graphene-based systems \cite{yuan-trilayer,abc-graphene,bernal} where the spin-orbit coupling responsible for pinning the $\v{d}$-vector is extremely weak. 

\begin{figure}
    \centering
    \includegraphics[width=85mm]{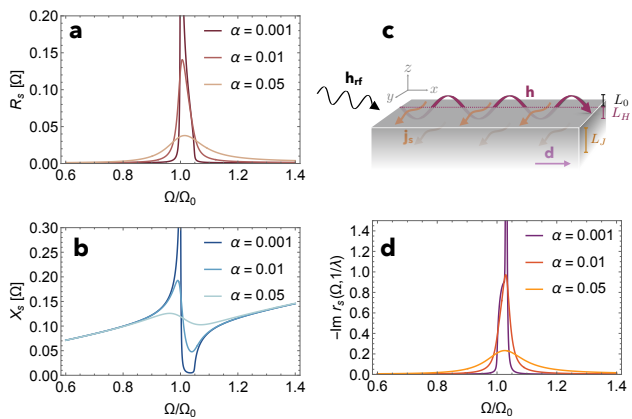}
    \caption{\textbf{Experimental signatures:} \textbf{a,b.} Surface resistance and reactance $Z_s = R_s - i X_s$ calculated within the two-fluid model. Both exhibit peaks at the bare Cooper pair spin wave frequency $\W_0$ and $X_s$ features a dip at the bulk dipolar mode frequency. \textbf{c.} Illustration of the surface mode in a semi-infinite sample with an in-plane $\v{d}$-vector. The magnetic field profile of the mode propagates in the plane of the interface, and decays over a length $L_H \sim \lambda$ inside the triplet superconductor and over a length $L_0$ in vacuum, while the supercurrent profile of the mode decays over a length $L_J$. \textbf{d.} Near-field reflection coefficient evaluated at $q = 1/\lambda$, which exhibits a peak at the surface mode frequency.}
    \label{exp-fig}
\end{figure}

\textit{Experimental detection --} Cooper pair spin waves contribute to experimentally observable electrodynamic response functions such as the surface impedance and reflectance of the triplet medium. We calculate these functions within a two-fluid model of electrodynamics where the charge response is described via the conductivity $\sigma(\W) = \sigma_n(\W) + i/(\lambda^2 \W)$. The first term describes the dissipative conductivity due to quasiparticles, while the second describes the kinetic inductance of the condensate. The complex surface impedance, which can be measured using cavity resonator techniques, is then given by $Z_s(\W) = \sqrt{\mu(\W)/\epsi(\W)}$, where $\epsi(\W) = 1 + i\sigma(\W)/\W$ is the dielectric function of the triplet medium (see the Supplemental Information for details). The surface resistance and reactance, $Z_s = R_s - iX_s$, are plotted in Fig. \ref{exp-fig}{\color{nrppurple}a,b} and exhibit a strong resonance at the bare Cooper pair spin wave frequency $\W_0$, as well as a weaker feature at the frequency of the bulk dipolar mode. 

The surface Cooper pair spin wave lies outside of the light cone (i.e. disperses with a velocity much slower than the speed of light), and thus must be excited by a near-field source, such as an antenna or microwave stripline \cite{magnonscattering,magnon-mu,sw-nv}. The same is true for dipolar spin waves in ferromagnets, and consequently near-field microwave spectroscopy techniques are well-developed \cite{sw-book,magnonics-iop,magnonics-review}. The appropriate response function to describe such a measurement is the near-field reflection coefficient $r_s(\W,q)$ \cite{sun-nearfield,ford-molecules} which is calculated in the Supplemental Information. This reflection coefficient is presented in Fig. \ref{exp-fig}{\color{nrppurple}d}, and features a peak at the surface mode frequency. This demonstrates that surface Cooper pair spin waves can be excited and detected much like other dipolar spin waves. For example, they can be excited by a microwave transmission line and detected either inductively by a second transmission line \cite{magnonics-iop,magnonics-review}, or electrically via the inverse spin Hall effect \cite{magnon-spintronics,spinhall}. 

Finally, recent advances in nitrogen-vacancy magnetometry \cite{amir-nv-review,uri-wte2} have enabled nitrogen-vacancy sensors to operate at the mK temperatures needed to access triplet superconductivity \cite{eth-nv}, allowing for the magnetic field profile of the surface modes to be directly imaged, just as Damon-Eshbach modes have recently been imaged in ferromagnets \cite{magnonscattering}. 

\textit{Conclusions -- } We have identified a class of magnetostatic modes in triplet superconductors and demonstrated their coupling to experimental probes. The detection of these modes in a given material would constitute strong evidence for spin triplet superconductivity, just as the detection of various collective modes in $^3$He proved crucial to identifying distinct superfluid phases \cite{3he-book,halperin-ch}. Our results complement prior works on orbital collective modes in multi-component superconductors \cite{power-1,power-2,Yip1992,214-1d,valentin-1,214-anisotropic,chiral-higgs,clappingmodes}, and tighten the analogies between unconventional superconductors, superfluid $^3$He, and magnetic materials.

\begin{acknowledgments}
The authors thank Bertrand Halperin, James Sauls, William Halperin, Andrew Zimmerman, Zhiyuan Sun, Yuan Cao, Uri Vool, Ilya Esterlis, Ian Hayes, Yonglong Xie, Manfred Sigrist, Mark Fischer, and Aharon Kapitulnik for fruitful discussions about this work. N.R.P. and J.B.C. acknowledge the hospitality of the ETH Z\"urich Institute for Theoretical Physics and the Max Planck Institute for the Structure and Dynamics of Matter (MPSD, Hamburg), where part of this work was completed. N.R.P., J.B.C., C.G.L.B., A.Y., and P.N. are supported by the Quantum Science Center (QSC), a National Quantum Information Science Research Center of  the  U.S.  Department  of  Energy  (DOE). N.R.P. is supported by the Army Research Office through an NDSEG fellowship. J.B.C. is an HQI Prize Postdoctoral Fellow and gratefully acknowledges support from the Harvard Quantum Initiative. 
V.M.G. is supported by the National Science Foundation under Grant No. DMR-2037158, the U.S. Army Research Office under Contract No. W911NF1310172, and the Simons Foundation.  A.Y. is partly supported by 
the Gordon and Betty Moore Foundation through Grant GBMF 9468 and by the National Science Foundation under Grant No. DMR-1708688. P.N. is grateful for the hospitality of the Max Planck Institute for the Structure and Dynamics of Matter where part of this work was completed supported by a Max Planck Sabbatical Award and a Bessel Research Award of the Alexander von Humboldt Foundation.
P.N. is a Moore Inventor Fellow and gratefully acknowledges support through Grant GBMF8048 from the Gordon and Betty Moore Foundation. E.D. acknowledges support from the ETH Z\"urich Institute for Theoretical Physics.
\end{acknowledgments}

\appendix 

\section{RPA theory for spin waves} \label{rpa-appendix}
To develop a microscopic weak coupling theory for Cooper pair spin waves, we begin with the BCS action for spin triplet pairing in imaginary time \cite{odd-w,eugene-spinwaves},
\begin{equation}
    S = \sum_k \bar{\psi}_{k\alpha} \big(-i\w_n + \xi_{\k} \big) \psi_{k\alpha} - \frac{1}{2}\sum_{q,i,\ell} g_i \, \bar{P}_q^{i,\ell} P_q^{i,\ell} \, ,
\end{equation}
where $\alpha = \uparrow,\downarrow$ is a spin index, $k = (i\w_n,\v{k})$ is a four-momentum with the fermionic Matsubara frequency $i\w_n$, $\xi_{\k}$ is the single-electron dispersion, and $g_i$ are coupling constants in each of the three spin triplet channels, $\hat{T}_i = i \hat{\sigma}_i \hat{\sigma}_2$, with $\hat{\sigma}_i$ the Pauli matrices in spin space. Finally, the Cooper pair operators (with center-of-mass momentum $q = (i\W_n,\v{q})$ and bosonic Matsubara frequency $i\W_n$) in each channel are $P_{q}^{i,\ell} = \sum_k \chi_{\k}^\ell T^{\alpha \beta}_i \, \psi_{-k,\alpha}\psi_{k+q,\beta}$ where $\chi_{\k}^\ell$ are orthogonal orbital form factors belonging to irreducible representations of the point group. Note that we have assumed that the coupling constants $g_i$ are independent of the orbital index, corresponding to pairing in a single (possibly multi-dimensional) orbital channel. 


To decouple the interaction term, we introduce Hubbard-Stratonovich fields $d_q^{i,\ell}$ in each channel, with the action $S_{\text{HS}} = \frac{1}{2}\sum_q g_i^{-1} \bar{d}_q^{i,\ell} d_q^{i,\ell}$. Shifting $d_q^{i,\ell} \goes d_q^{i,\ell} + g_i P_q^{i,\ell}$ and subsequently integrating out the electrons, we arrive at the effective action for the order parameter fields
\begin{equation} \label{eff-act-rpa}
    S = \frac{1}{2}\sum_{q,i,\ell} g_i^{-1} \,\bar{d}_q^{i,\ell} d_q^{i,\ell} - \frac{1}{2} \, \tr \log \check{\mathbb{G}}^{-1} \, ,
\end{equation}
where the inverse Gor'kov Green function is 
\begin{equation}
    \check{\mathbb{G}}^{-1}_{k,q} = \begin{pmatrix} \big(i\w_n - \xi_{\k} \big) \delta_{q,0} \hat{\id} & - \sum_{i,\ell} \chi_{\k}^\ell \hat{T}_i d^{i,\ell}_q \\[5pt] -\sum_{i,\ell} \chi_{\k}^\ell \hat{\bar{T}}_i \bar{d}^{i,\ell}_q & \big(i\w_n + \xi_{\k} \big) \delta_{q,0} \hat{\id}\end{pmatrix} \, .
\end{equation}
Throughout, we denote 2$\times$2 matrices in spin space with a hat, and 4$\times$4 matrices in spin $\otimes$ Nambu space with a check. Introducing the Pauli matrices in Nambu space, $\tau_i$, and $\tau^\pm = \frac{1}{2} \big(\tau_1 \pm i\tau_2\big)$, the saddle point equations are
\begin{equation}
    d_{q}^{i,\ell} = -g_i T \sum_{k} \chi_{\k}^{\ell} \, \tr \big[ \check{\mathbb{G}}_{k,q} \hat{\bar{T}}_i \, \tau^- \big] \, .
\end{equation}

For the sake of simplicity, we will now specialize to the case of $p$-wave pairing in two spatial dimensions, although our results can be readily generalized to other pairing symmetries. The two orbital form factors are $\chi_{\k}^{(x)} = \sqrt{2} \cos \phi_{\k}$ and $\chi_{\k}^{(y)} = \sqrt{2} \sin \phi_{\k}$, where  $\phi_{\k}$ is the angle around the Fermi surface. It is convenient to adopt a chiral basis \cite{clappingmodes}, with the form factors $\chi_{\k}^{(\pm)} = \frac{1}{\sqrt{2}}\big(\chi^{(x)}_{\k} \pm i\chi^{(y)}_{\k} \big) = \e^{\pm i \phi_{\k}}$ and the corresponding order parameters $d_{q}^{i,\pm}$. Finally, we will allow for an easy-axis spin anisotropy by taking $g_z > g_x = g_y \equiv g_\perp$. To compare to the main text, the dimensionless measure of the spin anisotropy is then
\begin{equation}
    \gamma_{\text{BCS}} = \frac{g_\perp^{-1} - g_z^{-1}}{\nu} \, ,
\end{equation}
where $\nu$ is the density of states at the Fermi level. 

We expand around the homogeneous saddle point where $\v{d}^+_{q} = \Delta_0 \uv{z}$ and $\v{d}^-_q = 0$, corresponding to a $p+ip$ state. The saddle point equation then simply becomes
\begin{equation}
    g_z^{-1} = - T \sum_k \frac{1}{(i\w_n)^2 - E_{\k}^2} \, ,
\end{equation}
where $E_{\k}^2 = \xi_{\k}^2 + \Delta_0^2$. We can then parameterize the collective modes of the system as
\begin{align}
    \v{d}_q^+ &= \e^{i\varphi_q} \, \big[(\Delta_0 + h_q) \uv{z} + \v{a}_q + i \v{b}_q \big] \\[5pt]
    \v{d}_q^- &= \e^{i\varphi_q} \, \big[ \v{c'}_q + i \v{c''}_q  \big] \, ,
\end{align}
where $\varphi$ is the global phase mode, $h$ is the Higgs mode, and $\v{a} = (a_x,a_y,0)$ and $\v{b} = (b_x,b_y,0)$ are the transverse Cooper pair spin waves. Additionally, there are six orbital ``clapping modes'' in the $p-ip$ channel \cite{214-modes,clappingmodes}, which are all degenerate at frequency $\W_{\text{cm}} = \sqrt{2} \Delta_0$ and decouple from the other collective modes. Further, the phase and Higgs modes also decouple from the spin waves, so we may restrict our attention to the transverse sector, in which case the fluctuating Gor'kov Green function becomes 
\begin{equation} \label{vertices}
    \check{\mathbb{G}}^{-1}_{k,q} = \check{\mathbb{G}}^{-1}_{k} - \check{\Lambda}^{\v{a}}_{k,q} - \check{\Lambda}^{\v{b}}_{k,q} \, ,
\end{equation}
where the mean-field Green function and spin waves vertices are
\begin{align}
    \check{\mathbb{G}}^{-1}_k &= i\w_n - \xi_{\k}\tau_3 - \Delta_0 \big( \chi_{\k}^+ \tau^+ + \chi_{\k}^- \tau^- \big) \hat{T}_z \\[4pt]
    \check{\Lambda}^{a_i}_{kq} &= \big( \chi_{\k}^+ \tau^+ \hat{T}_i + \chi_{\k}^- \tau^- \hat{\bar{T}}_i \big) a_{i,q} \\[4pt]
    \check{\Lambda}^{b_i}_{kq} &= i \big( \chi_{\k}^+ \tau^+ \hat{T}_i - \chi_{\k}^- \tau^- \hat{\bar{T}}_i \big) b_{i,q} \, .
\end{align}
Expanding the functional logarithm in (\ref{eff-act-rpa}) to quadratic order in the spin waves, one finds 
\begin{equation}
    S = - \frac{1}{2}\sum_q \, \Big( D_{\v{a}}^{-1}(q) \v{a}_{-q} \cdot\v{a}_q + D_{\v{b}}^{-1}(q) \v{b}_{-q} \cdot \v{b}_q \Big)\, ,
\end{equation}
where, at zero temperature, the retarded propagators for the spin waves are
\begin{align}
    D_{\v{a}}^{-1}(q) &= \big( \W^2 - v_F^2 \v{q}^2/2 \big) f(\W) - (g_\perp^{-1} - g_z^{-1}) \\[4pt]
    D_{\v{b}}^{-1}(q) &= \big(\W^2 - 4\Delta_0^2 - v_F^2 \v{q}^2/2 \big) f(\W) - (g_\perp^{-1} - g_z^{-1})
\end{align}
where $v_F$ is the Fermi velocity and 
\begin{equation}
    f(\W) = \frac{\nu}{2\Delta_0 \W}\frac{\sin^{-1}(\W/2\Delta_0)}{\sqrt{1- (\W/2\Delta_0)^2}} \, ,
\end{equation}
is the Tsuneto condensate response function \cite{tsuneto}, where $\nu$ is the density of states at the Fermi level. Note that for a fully spin-rotation symmetric system, $\v{a}$ is a linearly dispersing Goldstone mode, while $\v{b}$ is overdamped and resides at the gap edge. The introduction of spin anisotropy gaps the $\v{a}$ mode. For weak anisotropy, $\W_0 \ll 2\Delta$ and we can approximate the spin wave gap to be
\begin{equation}
    \W_0^2 = \pfrac{g_\perp^{-1} - g_z^{-1}}{\nu} 4 \Delta_0^2 = 4\gamma_{\text{BCS}} \Delta_0^2 \, .
\end{equation}
For weak spin anisotropy, we can use the weak-coupling relation $T_c^{(i)} \sim \e^{-1/g_i\nu}$ to relate the difference in coupling constants to the transition temperatures in each channel according to $(g_\perp^{-1} - g_z^{-1})/\nu \approx (T_c^{(z)} - T_c^{(\perp)})/T_c^{(z)} \equiv \delta T_c/T_c$. 

We note that in general, there is also a linear-in-frequency coupling between the $\v{a}$ and $\v{b}$ modes which is proportional to the degree of particle-hole asymmetry and scales as $\Delta_0/\epsi_F$. However, in the weak coupling limit, we may adopt the quasi-classical approximation  $\Delta_0/\epsi_F \goes 0$, and neglect this coupling. 

To study the interaction between the spin waves and a fluctuating magnetic field $\v{h}$, we may include a Zeeman coupling between the electrons and $\v{h}$. This is accomplished by including an additional vertex in (\ref{vertices}), $\check{\Lambda}^{\v{h}}_{q} = \mu_B \v{h}_{q} \cdot \check{\v{s}}$, where $\mu_B$ is the Bohr magneton and the electron spin operator in Nambu space is
\begin{equation}
    \check{\v{s}} = \pfrac{\tau_3 + 1}{2} \, \v{\hat{\sigma}} + \pfrac{\tau_3 - 1}{2} \, \v{\hat{\bar{\sigma}}} \, .
\end{equation}
Then, expanding (\ref{eff-act-rpa}) to quadratic order in $\v{h}$ generates a linear-in-frequency coupling between the magnetic field and $\v{a}$. In contrast, the coupling between $\v{b}$ and $\v{h}$ is proportional to the particle-hole asymmetry $\sim \Delta_0/\epsi_F$, and vanishes in the quasi-classical approximation. At zero temperature and in real time, the coupling between $\v{a}$ and $\v{h}$ is
\begin{equation}
    \delta S = - 2i \mu_B \Delta_0 \sum_q \W f(\W) \, \uv{z} \cdot \left( \v{a}_{-q} \times \v{h}_q \right) \, .
\end{equation}
Integrating out the spin waves and including the renormalization due to the condensate itself, we can identify the renormalized ($\v{q} = 0$) dynamic susceptibility to be $\chi_{xx} = \chi_{yy} = \chi_\perp$, with 
\begin{equation}
    \chi_\perp(\W) = 4\mu_B^2 \Delta_0^2 f(\W) \left[ \frac{g_\perp^{-1} - g_z^{-1}}{(g_\perp^{-1} - g_z^{-1})-\W^2 f(\W)} \right] \, .
\end{equation}
In the limit of weak anisotropy, this takes the simple form 
\begin{equation}
    \chi_{\perp}(\W) = \chi_0 \, \frac{\W_0^2}{\W_0^2 - \W^2} \, ,
\end{equation}
where $\chi_0 = \nu \mu_B^2$ is the normal state Pauli paramagnetic susceptibility. Crucially, this susceptibility is of the same form as the undamped susceptibility derived in the main text using the time-dependent Ginzburg-Landau theory.

\section{Surface modes} \label{magnetostatics-appendix}
\begin{figure}
    \centering
    \includegraphics[width=65mm]{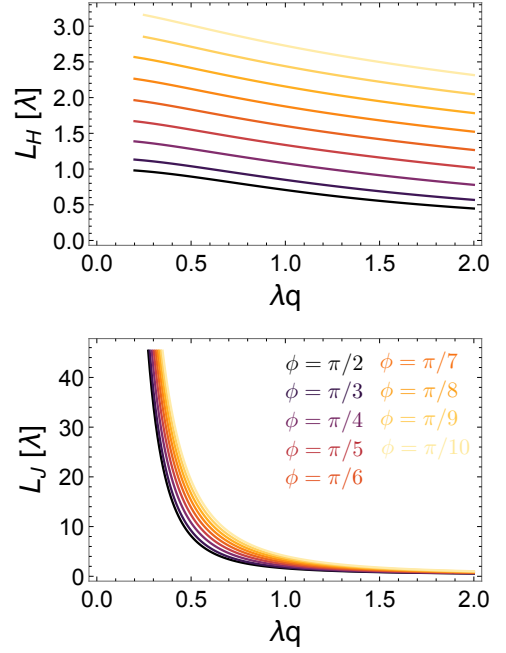}
    \caption{Decay lengths for the magnetic field ($L_H$) and supercurrent ($L_J$) within the magnetostatic approximation.}
    \label{decaylengths-fig}
\end{figure}

In this section, we present a derivation of the surface Cooper pair spin wave dispersion within the magnetostatic approximation, which amounts to taking $c \goes \infty$. We consider a triplet superconductor occupying the half-space $z <0$, and vacuum for $z >0$. Outside the triplet superconductor, the quasi-magnetostatic Maxwell equations are simply $\nabla \cdot \v{b} = 0$ and $\nabla \times \v{h} = 0$, with the trivial constitutive relation $\v{b} = \v{h}$. The solution to these equations that is exponentially localized to the $z = 0$ surface is 
\begin{equation}
    \v{h}_{\text{out}} = \tilde{h}_0 \, \begin{pmatrix} q_x/q_y \\ 1 \\ i \kappa_0/q_y  \end{pmatrix} \, \e^{-\kappa_0 z} \, \e^{iq_x x+ iq_y y} \, ,
\end{equation}
with the decay length 
\begin{equation}
    \kappa_0 = \sqrt{q_x^2 + q_y^2} \, .
\end{equation}
In the triplet superconductor, we take the $\v{d}$-vector to lie in the $\uv{x}$ direction so that the magnetic permeability tensor is $\underline{\mu} = \text{diag}(1,\mu,\mu)$. The quasi-magnetostatic Maxwell-London equations are 
\begin{subequations}
\begin{align}
    \nabla \cdot \v{b} &= 0 \\
    \nabla \times \v{h} &= \v{j}_s \\
    \nabla \cdot \v{j}_s &= 0 \\
    \nabla \times \v{j}_s &= - \lambda^{-2}\v{b} \, ,
\end{align}
\end{subequations}
with $\v{b} = \underline{\mu}\v{h}$. The solution to these equations localized to the $z=0$ surface is
\begin{align}
    \v{h}_{\text{in}} &= \begin{pmatrix} h_0 \e^{\kappa_H z} \\[5pt] \frac{\lambda^2}{\mu+\lambda^2 q_x^2} \big(q_x q_y h_0 \e^{\kappa_H z} - \kappa_J j_0 \e^{\kappa_J z} \big) \\[5pt] \frac{\lambda^2}{\mu + \lambda^2 q_x^2} \big(iq_y j_0 \e^{\kappa_J z} - iq_x \kappa_H h_0 \e^{\kappa_H z} \big)
     \end{pmatrix} \e^{iq_x x + iq_y y} \\[6pt]
     \v{j}_{s} &= \begin{pmatrix} j_0 \e^{\kappa_J z} \\[5pt] \frac{1}{\mu + \lambda^2 q_x^2}\big( \lambda^2 q_x q_y j_0 \e^{\kappa_J z} +\mu \kappa_H h_0 \e^{\kappa_H z}\big) \\[5pt] \frac{-1}{\mu + \lambda^2 q_x^2} \big( iq_y \mu h_0 \e^{\kappa_H z} + \lambda^2 q_x \kappa_J j_0 \e^{\kappa_J z} \big) \end{pmatrix} \e^{iq_x x + iq_y y}
\end{align}
where $h_0$ and $j_0$ are constants, and the decay lengths are
\begin{subequations}
\begin{align}
   1/L_J &= \kappa_J = \sqrt{\mu \lambda^{-2} + q_x^2 + q_y^2} \\[5pt]
   1/L_H &= \kappa_H = \sqrt{ \lambda^{-2} + \mu^{-1} q_x^2 + q_y^2} \, .
\end{align}
\end{subequations}
These decay lengths are plotted in Fig. \ref{decaylengths-fig} as a function of the angle $\cos \phi = \v{q}\cdot \v{d}$ between the propogation direction and the $\v{d}$-vector.

We now impose boundary conditions. First, we require that $j_{s,z}(z=0) = 0$ to ensure supercurrent conservation. Enforcing this condition requires that
\begin{equation}
    h_0 = - \frac{\lambda^2 q_x \kappa_J}{q_y \mu} \, j_0 \, .
\end{equation}
Next, we require that the normal component of the magnetic induction be continuous across the interface, $\mu h_{\text{in},z}(z=0) = h_{\text{out},z}(z=0)$. This requires that
\begin{equation}
    \tilde{h}_0 = \frac{q_y^2 \mu + \lambda^2 q_x^2 \kappa_J \kappa_H}{\kappa_0 (\mu + \lambda^2 q_x^2 )} \, \lambda^2 j_0 \, .
\end{equation}
Finally, we require that the tangential components of the magnetic field be continuous across the interface. Both $h_x$ and $h_y$ are continuous across the interface provided that 
\begin{equation}
    \frac{\kappa_J}{\mu} + \frac{\mu q_y^2 + \lambda^2 q_x^2 \kappa_J \kappa_H}{\sqrt{q_x^2 + q_y^2}\big(\mu + \lambda^2 q_x^2 \big) } = 0 
\end{equation}
is satisfied. The dispersion of the surface spin waves is determined implicitly from this equation by solving for $\W(\q)$.

\section{Two-fluid model}
In this section, we relax the magnetostatic approximation and calculate the full electrodynamic response of the surface modes within a two-fluid model of electrodynamics that allows us to include the effects of quasiparticles in addition to the superfluid response. Our analysis is based on classical electrodynamics, starting from Maxwell's equations
\begin{equation}
    \begin{split}
        \nabla \cdot \v{E} &= \rho \\
        \nabla \cdot \v{B} &= 0 \\ 
        \nabla \times \v{E} &= i\W \v{B} \\
        \nabla \times \v{H} &= \v{J} -i\W \v{D}  
    \end{split}
\end{equation}
For clarity, we keep factors of the electromagnetic constants $\mu_0$, $\epsi_0$, and $c$ explicit in this section. We then describe the response of the triplet medium via the consistutive relations
\begin{equation}
    \begin{split}
        \v{B} &= \underline{\mu} \v{H} \\
        \v{D} &= \underline{\epsi} \v{E} \, ,
    \end{split}
\end{equation}
Taking the $\v{d}$-vector to lie in the $\uv{y}$ direction, the magnetic permeability $\underline{\mu} = \text{diag}(\mu,1,\mu)$ is anisotropic and features a resonance in the components of $\underline{\mu}$ transverse to $\v{d}$, reflecting the condensate spin wave mode,
\begin{equation}
    \mu(\W) = \mu_0 \left(1 + \chi_0 \, \frac{\W_0^2}{\W_0^2 - \W^2} \right) \, ,
\end{equation}
where $\W_0$ is the bare spin wave frequency, and the dimensionless constant $\chi_0$ determines the static relative permeability $\mu_r(0) = \mu(0)/\mu_0$. For simplicity, we neglect spin wave damping for the time being.

The charge response is specified by the ac conductivity, which relates $\v{J} = \sigma \v{E}$. We take the conductivity to be of the two-fluid form, 
\begin{equation}
    \sigma(\W) = \sigma_n(\W) + \frac{i \Lambda}{\W} \equiv \sigma_1(\W) + i\sigma_2(\W) \, ,
\end{equation}
where $\sigma_n$ is the dissipative normal fluid conductivity, and the second term describes the kinetic inductance of the condensate. Combined with Maxwell's equations, this conductivity reproduces the longitudinal and transverse London response, with the superfluid weight $\Lambda = 1/(\mu_0 \lambda^2)$ expressed in terms of the penetration depth, $\lambda$. 

For simplicity, we take the bare dielectric constant of the triplet medium to be $\underline{\epsi} = \epsi_0 \underline{\id}$, which is dressed by the conductivity to give the effective dielectric function
\begin{equation}
    \epsi(\W) = \epsi_0 \left(1 + \frac{i\sigma(\W)}{ \epsi_0 \W}\right) \, .
\end{equation}
In the simplest case $\sigma_n = 0$, this becomes 
\begin{equation}
\epsi(\W) = \epsi_0 \left(1 - \frac{\W_{\text{pl}}^2}{\W^2} \right) \, ,
\end{equation}
where $\W_{\text{pl}} = c/\lambda$ is the superfluid plasma frequency. Restoring $\sigma_n \neq 0$ then broadens the plasmon resonance. 

\begin{figure}
    \centering
    \includegraphics[width=80mm]{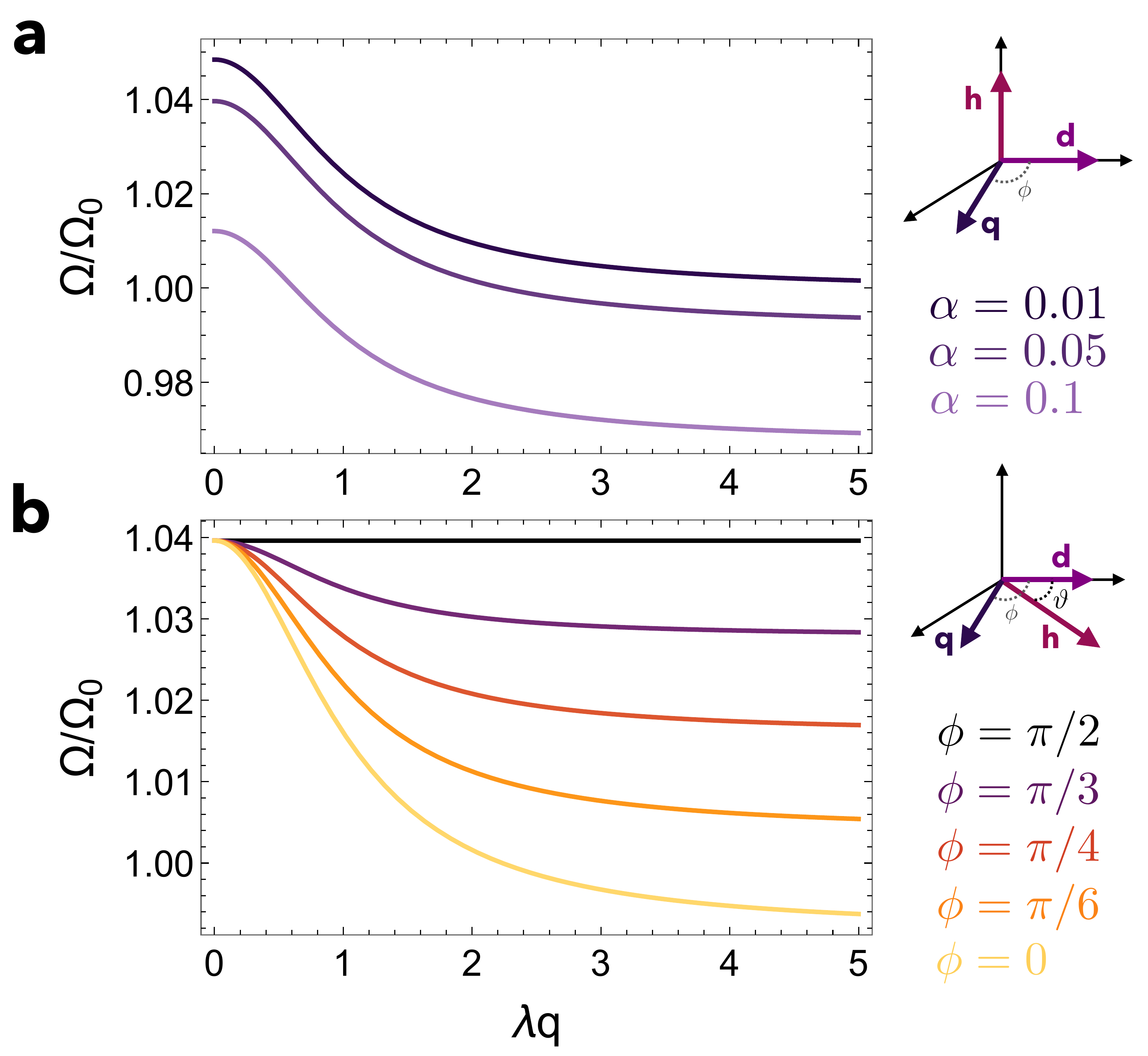}
    \caption{\textbf{a.} Dispersion of the transverse bulk mode for several values of the damping parameter $\alpha$, taking $\Gamma/\W_b = 0.05$ and $\chi_0 = 0.1$. \textbf{b.} Dispersion of the partially-longitudinal bulk mode for several directions of propagation, with $\alpha = 0.05$. Diagrams illustrating the orientation of $\v{q}, \v{h}$, and $\v{d}$ for each mode are shown next to the corresponding plot. In panel $\textbf{b}$, the polarization $\vartheta$ of $\v{h}$ with respect to $\v{d}$ is given by $\tan \vartheta = 2(\lambda^2 q^2 + \sin^2 \phi)/\sin 2\phi$. }
    \label{inf-medium}
\end{figure}

Using the constitutive relations outlined above, and the continuity equation $-i\W \rho + \nabla \cdot \v{J} = 0$ to eliminate $\rho$, we arrive at a set of closed equations for $\v{E}$ and $\v{H}$,
\begin{equation}
    \begin{split}
        \epsi \, \nabla \cdot \v{E} &= 0 \\
        \nabla \cdot \underline{\mu} \v{H} &= 0 \\
        \nabla \times \v{E} &= i\W \underline{\mu} \v{H} \\
        \nabla \times \v{H} &= -i\W \epsi \v{E} 
    \end{split}
\end{equation}
The last two equations can be combined to give a wave equation for $\v{H}$,
\begin{equation}
    \left(\W^2 \epsi \mu_{ij} + \delta_{ij} \del_k^2 - \del_i \del_j \right) H_j = 0 \, ,
\end{equation}
while the first equation requires that the longitudinal component of $\v{E}$ vanishes for $\W < \W_{\text{pl}}$, and the transverse component is determined from $\v{H}$ as
\begin{equation}
    \v{E} = \frac{i}{\W \epsi} \nabla \times \v{H} \, .
\end{equation}
In an infinite translationally invariant system, there are two branches of solutions to these equations, which correspond to two bulk dipolar spin wave modes. In the magnetostatic limit, $c \goes \infty$, the dispersions $\W(\v{q})$ of each mode are determined implicitly through the conditions 
\begin{align}
   0 &= \mu(\W) + \lambda^2 q^2  \\
    0 &= \left(1 + \lambda^2 q^2 \sin^2 \phi \right) \mu(\W) + \lambda^2 q^2 \cos^2 \phi  \, 
\end{align}
where $\phi$ is the angle between $\v{q}$ and $\v{d}$. The dispersions of these bulk dipolar modes are plotted in Fig \ref{inf-medium}.

In particular, note that for $q = 0$, the two branches become degenerate, with the mode frequency determined by the zero of the permeability, $\mu(\W) =0$. This is in contrast to the bare spin wave, which corresponds to a pole in $\mu(\W_0) \goes \infty$. 

We now wish to study solutions of the system above for a semi-infinite triplet superconductor occupying the half-space $z > 0$, with vacuum below ($z < 0$). We again take the $\v{d}$ vector to lie along the $\uv{y}$ axis, perpendicular to the interface normal, and consider the simplest trial solution which propagates in the $\uv{x}$ direction, perpendicular to $\uv{d}$. That is, we take $\v{H}(\v{r}) = \v{H}(\v{z}) \e^{iqx}$, where $q$ is the in-plane wavevector.

We further take the magnetic field to be polarized in the $xz$ plane, which leads to two coupled equations for $H_x$ and $H_z$,
\begin{align}
    \frac{\W^2 \epsi \mu}{\w^2 \epsi \mu - q^2} \left( \W^2 \epsi \mu - q^2 + \del_z^2 \right) H_x = 0 \\
    H_z = \frac{iq}{\W^2 \epsi \mu - q^2} \, \del_z H_x \, .
\end{align}
These equations are satisfied by defining the inverse decay length 
\begin{equation}
    \kappa_\perp = \sqrt{q^2 - \W^2 \epsi \mu} 
\end{equation}
such that the magnetic field profile inside the triplet medium is 
\begin{equation}
    \v{H}_{\text{in}} = H_{\text{in}} \, \e^{iqx} \, \e^{-\kappa_\perp z} \, \begin{pmatrix} 1 \\ 0 \\  iq/\kappa_\perp \end{pmatrix} \, ,
\end{equation}
and the electric field is then determined to be 
\begin{equation}
    \v{E}_{\text{in}} = \frac{i\W \mu}{\kappa_\perp} \, H_{\text{in}} \, \e^{iqx} \, \e^{-\kappa_\perp z} \, \begin{pmatrix} 0 \\ 1 \\ 0 \end{pmatrix} \, .
\end{equation}
Outside the medium, one can repeat this procedure, where one finds a field localized to the $xy$ plane takes the form 
\begin{align}
    \v{H}_{\text{out}} &= H_{\text{out}} \e^{iqx} \, \e^{\kappa_0 z} \begin{pmatrix} 1 \\0 \\ -iq/\kappa_0 \end{pmatrix} \\[6pt]
    \v{E}_{\text{out}} &= -\frac{i\W \mu_0}{\kappa_0} H_{\text{out}}  \e^{iqx} \, \e^{\kappa_0 z} \begin{pmatrix} 0 \\ 1 \\ 0 \end{pmatrix} \, ,
\end{align}
where the inverse decay length in vacuum is 
\begin{equation} 
    \kappa_0 = \sqrt{q^2 - \W^2/c^2}.
\end{equation}
We now enforce the continuity of $H_x$, $E_y$, and $B_z$ across the $z = 0$ interface. These boundary conditions can only be satisfied at a particular frequency that is implicitly determined from
\begin{equation} \label{de-disp-eq}
    1 + \frac{\mu_r(\W) \kappa_0}{\kappa_\perp} = 0 \, .
\end{equation}
This determines the dispersion of the surface mode, and reduces to our prior magnetostatic result in the limit $c \goes \infty$. 

\begin{figure}
    \centering
    \includegraphics[width=65mm]{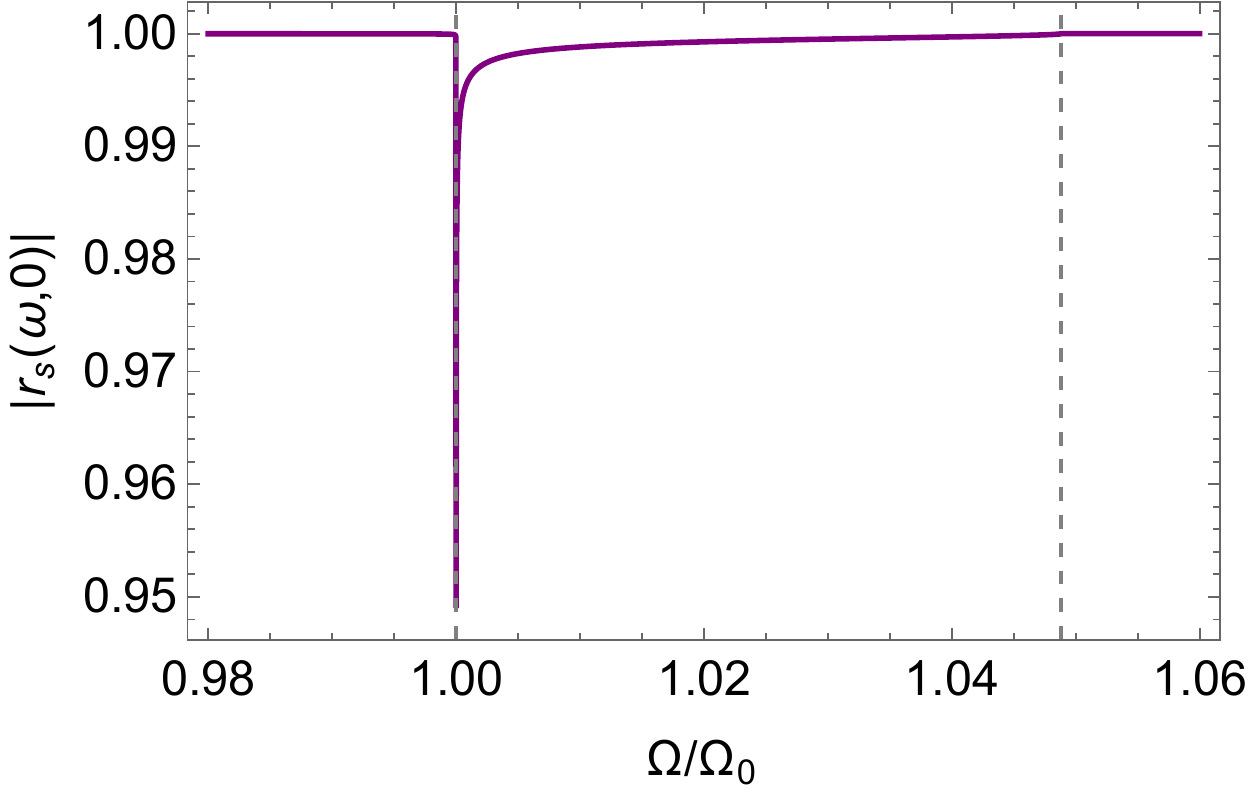}
    \caption{Far field reflection coefficient $r_s$ evaluated at zero in-plane momentum. The magnitude of the reflection coefficient is one below the plasma frequency, and features a dip at the bare spin wave frequency $\W = \W_0$, with non-zero transmission up to the top of the bulk dipolar spin wave band, determined from $\mu(\W) = 0$ and marked by the second dashed line in the bottom panel.}
    \label{fresnel-fig}
\end{figure}

\section{Surface impedance}
Next, we consider the reflection and transmission of plane waves off of a triplet superconductor in the same geometry. We consider $s$-polarized incident and reflected waves, with the wave-vectors
\begin{equation}
    \v{k}_I = \frac{\W}{c} \begin{pmatrix}\sin \theta \\ 0 \\ \cos \theta  \end{pmatrix} \, , \quad \v{k}_R = \frac{\W}{c} \begin{pmatrix}\sin \theta \\ 0 \\ -\cos \theta  \end{pmatrix}
\end{equation}
and fields 
\begin{align*}
    \v{E}_I &= Z_0 H_I \e^{i\v{k}_I\cdot \v{r}} \begin{pmatrix} 0 \\ 1 \\ 0 \end{pmatrix} 
   \, , \quad  \v{H}_I = H_I \e^{i\v{k}_I\cdot \v{r}} \begin{pmatrix} -\cos \theta \\ 0 \\ \sin \theta \end{pmatrix} \\[6pt]
   \v{E}_R &= Z_0 H_R \e^{i\v{k}_R\cdot \v{r}} \begin{pmatrix} 0 \\ 1 \\ 0 \end{pmatrix} 
   \, , \quad  \v{H}_R = H_R \e^{i\v{k}_R\cdot \v{r}} \begin{pmatrix} \cos \theta \\ 0 \\ \sin \theta \end{pmatrix} \, ,
\end{align*}
where $Z_0 = \sqrt{\mu_0/\epsi_0}$ is the vacuum impedance. We take the wave transmitted into the triplet medium to be of the form studied above, 
\begin{equation}
\begin{split} \label{ht-de}
    \v{H}_T &= H_T \e^{iqx - \kappa_\perp z} \begin{pmatrix} 1 \\ 0 \\ iq/\kappa_\perp \end{pmatrix} \\[5pt]
    \v{E}_T &= \frac{i\W \mu}{\kappa_{\perp}} H_T \e^{iqx-\kappa_\perp z} \begin{pmatrix} 0 \\ 1 \\ 0  \end{pmatrix} \, .
    \end{split}
\end{equation}

Conservation of in-plane momentum requires that $q = (\W/c) \sin \theta$, and imposing that $E_y$, $H_x$, and $B_z$ be continuous at $z = 0$ leads to the equations 
\begin{align}
    H_R + H_I &= \frac{i\W \mu}{Z_0 \kappa_\perp}\, H_T \\
    H_R - H_I &= \frac{\W/c}{\sqrt{\W^2/c^2 - q^2}} \, H_T \, .
\end{align}
Defining $k_0 = \sqrt{\W^2/c^2 - q^2}$, we can solve these equations for the  reflection coefficient $r_s = H_R/H_I$,
\begin{equation}
    r_s = \frac{ik_0 \mu_r + \kappa_\perp}{ik_0 \mu_r - \kappa_\perp} 
\end{equation}
We may also identify the surface impedance of the triplet medium as $Z_s = -E_y/H_x|_{z = 0^+}$ \cite{ford-molecules}, which gives
\begin{equation}
    Z_s = -\frac{i\w \mu}{\kappa_\perp} \, .
\end{equation}
For normal incidence with $q=0$, the surface impedance and reflection coefficient reduce to their usual textbook forms \cite{ll-v6},
\begin{align}
    r_s &= \frac{Z_s - Z_0}{Z_s + Z_0} \, ,\\[5pt]
    Z_s &= \sqrt{\frac{\mu(\w)}{\epsi(\w)}} \, .
\end{align}

\begin{figure}
    \centering
    \includegraphics[width=80mm]{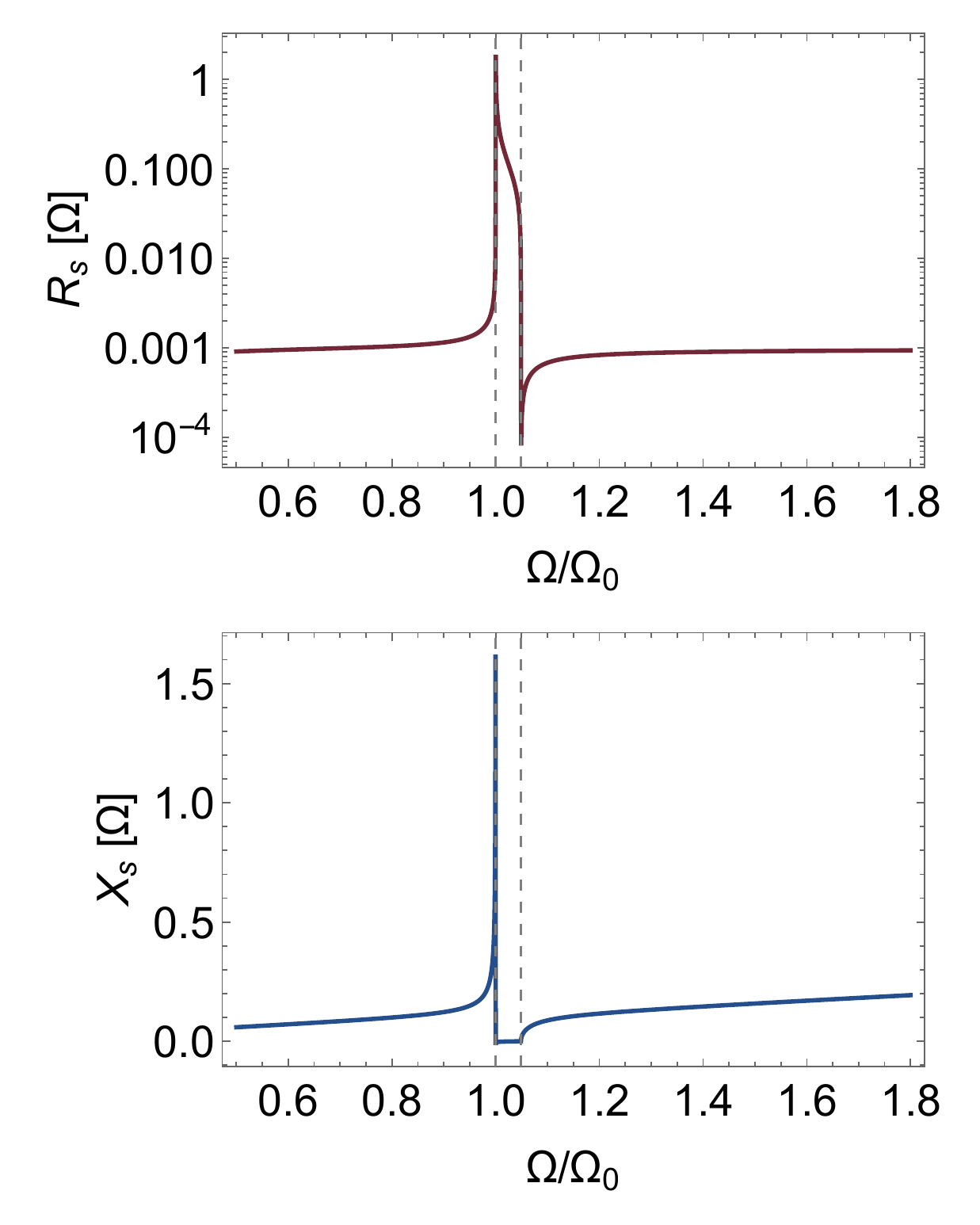}
    \caption{Surface resistance and reactance, $Z_s = R_s - iX_s$, near the collective mode frequency. Both display resonant features near the bare spin wave and bulk dipolar mode frequencies.}
    \label{zs-fig}
\end{figure}

Below the plasma frequency, $|r_s| = 1$ and light is fully reflected, except near the frequency of the spin collective modes. There is a dip peaked at the bare spin wave frequency $\W = \W_0$, with weight that extends up until the bulk dipolar spin wave frequency, determined from $\mu(\W) = 0$. This is because for fixed in-plane momentum, the bulk modes constitute a continuum as the out-of-plane momentum is varied.

Turning to the surface impedance, one identifies the real and imaginary parts to be the surface resistance and reactance, $Z_s = R_s - iX_s$, with the sign of $X_s$ chosen so that $X_s \sim \W$ follows the usual convention for an inductance ($Z \sim -i\W L$). Away from the spin wave resonances, we can approximate $\mu \approx \mu_0$, and expand $Z_s$ for $\W/\W_{\text{pl}} \ll 1$ and $\sigma_1/\sigma_2 \ll 1$, finding the usual behavior \cite{mw-sc-book}
\begin{align}
    R_s &\approx \frac{\mu_0^2}{2} \sigma_1(\W) \, \lambda^3 \, \W^2 \\
    X_s &\approx \mu_0 \lambda \, \W
\end{align}
which includes the effects of both dissipation through quasiparticles as well as the inductive condensate response. 

Near the spin wave resonance, we see that $Z_s \sim \sqrt{\mu(\W)}$, and thus will display resonant behavior at the poles and zeroes of $\mu(\W)$. As shown in Fig. \ref{zs-fig}, the bare spin wave manifests as a peak in the surface impedance (since it is a pole of $\mu(\W)$), whereas the bulk dipolar mode manifests as a dip (since it is a zero of $\mu(\W)$). 

Thus, we have shown that both the bare spin wave mode and the bulk dipolar mode can be detected via measurements of the surface impedance or reflectance. The surface mode, however, resides outside of the light cone and cannot couple to far field radiation. In light of this, in the next section we consider the near field response of the system, including the contribution of the surface spin wave mode.

\section{Near field response}

\begin{figure}
    \centering
    \includegraphics[width=70mm]{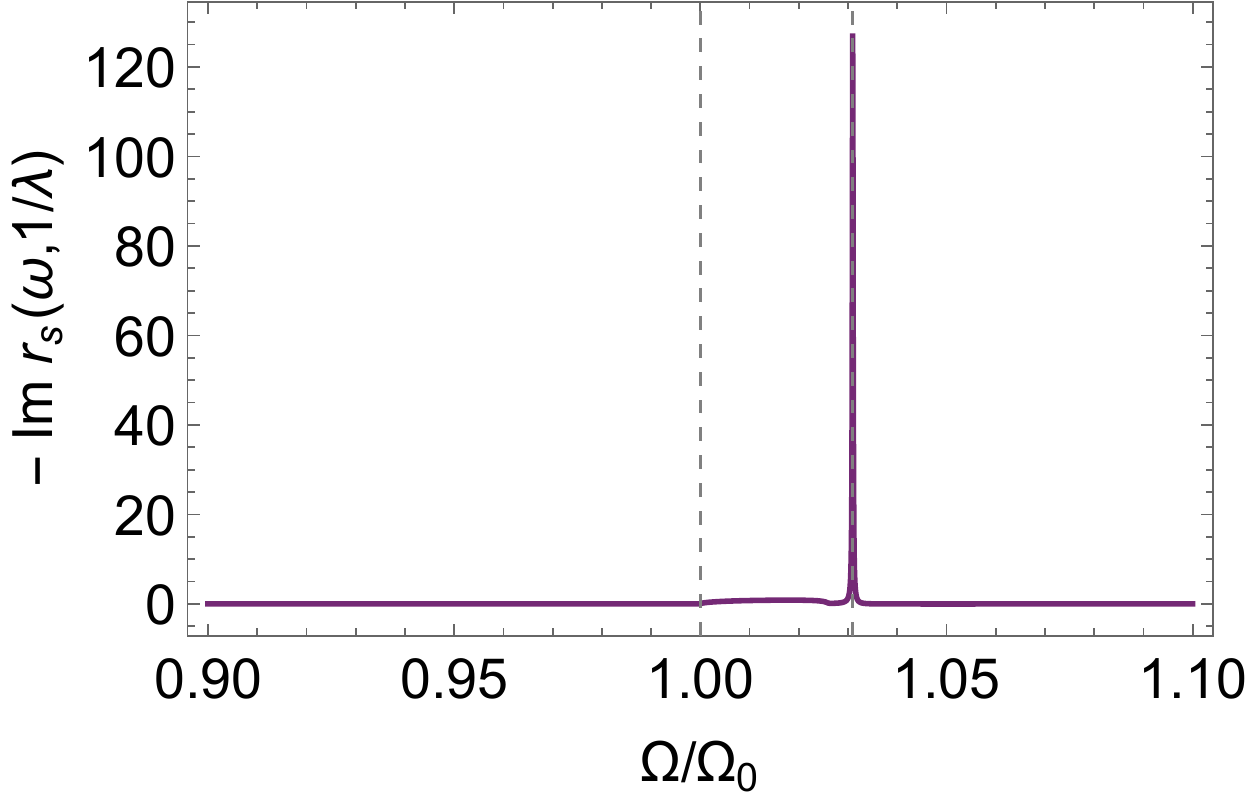}
    \caption[width=80mm]{Imaginary part of the near field reflection coefficient $r_s$ evaluated at in-plane momentum $q = 1/\lambda$. The dashed lines indicate the bare spin wave frequency and the surface mode frequency.}
    \label{near-field-fig}
\end{figure}

As the simplest model of a near field source, we study the reflection of an evanescent wave emanating from $z \goes -\infty$ and impinging on the triplet superconductor surface at $z = 0$. That is, we take the incident wave to be 
\begin{equation}
\begin{split}
      \v{H}_I  &= H_I \e^{iqx - \kappa_0 z} \begin{pmatrix} 1 \\ 0 \\ iq/\kappa_0 \end{pmatrix} \\[5pt]
     \v{E}_I &= \frac{i\mu_0 \W}{\kappa_0} H_I \e^{iqx - \kappa_0 z} \begin{pmatrix} 0 \\ 1 \\ 0 \end{pmatrix}
\end{split}
\end{equation}
where the decay length in vacuum is 
\begin{equation}
    \kappa_0 = \sqrt{q^2 - (\W/c)^2} \, .
\end{equation}
The reflected wave is then also evanescent, emanating from the $z = 0$ surface, 
\begin{equation}
\begin{split}
      \v{H}_R  &= H_R \e^{iqx + \kappa_0 z} \begin{pmatrix} 1 \\ 0 \\ -iq/\kappa_0 \end{pmatrix} \\[5pt]
      \v{E}_I &= -\frac{i\mu_0 \W}{\kappa_0} H_R \e^{iqx - \kappa_0 z} \begin{pmatrix} 0 \\ 1 \\ 0 \end{pmatrix} \, ,
\end{split}
\end{equation}
and the wave transmitted into the triplet medium is of the same form as (\ref{ht-de}). Again matching $H_x$, $E_y$, and $B_z$ at the $z = 0$ interface, we find the near field reflection coefficient to be 
\begin{equation}
    r_s = \frac{1 - \mu_r \kappa_0/\kappa_\perp}{1 + \mu_r \kappa_0/\kappa_\perp} \, .
\end{equation}
Comparing to (\ref{de-disp-eq}), we see that $r_s$ has a pole at the Damon-Eshbach mode frequency, manifesting as a strong peak seen in Fig. \ref{near-field-fig}.

\section{Spin wave damping}
Finally, we can re-incorporate the effects of spin wave damping within the time-dependent Ginzburg-Landau formalism. In this case, the transverse spin susceptibility is 
\begin{equation}
    \chi(\W) = \chi_0 \, \frac{\W_{\text{b}} (\Gamma - i \alpha \W)}{(\W_{\text{b}} - i\alpha \W)(\Gamma - i\alpha \W) - \W^2} \, ,
\end{equation}
where $\Gamma$ is the $\v{d}$-vector pinning energy, $\W_{\text{b}} \approx 2\Delta_0$ is the gap for non-unitary spin fluctuations, and $\alpha$ is the dimensionless spin wave damping parameter. We plot the surface impedance and near field reflection coefficients for several values of the damping parameter in Fig. 3 of the main text.

\section{Parameter values}
For all of the numerically generated plots above and in the main text, we use parameter values which are roughly appropriate for UPt$_3$. The superconducting gap is $2\Delta_0/2 \pi \approx 36 \; \text{GHz}$, and the $\v{d}$-vector pinning energy is approximately $\Gamma/2 \pi \approx 6 \; \text{GHz}$, as inferred from NMR experiments \cite{upt3-pinning,gannon-upt3}. This sets the spin wave frequency to be $\W_0/2\pi = \sqrt{\Gamma (2\Delta_0)} \approx 14 \; \text{GHz}$. 

We take $\lambda = 1 \; \mu\text{m}$ \cite{Gannon_2015,upt3-optics1,upt3-rmp} (which sets $\W_{\text{pl}} = 300 \; \text{THz}$) and $\chi_0 = 0.1$. For the normal conductivity, we use the typical Drude expression
\begin{equation}
    \sigma_n(\W) = \frac{D_n \, \tau}{1 - i \W \tau} \, , 
\end{equation}
where, to minimize the number of parameters, we take the Drude weight to be $D_n = f_n \epsi_0 \W_{\text{pl}}^2$, where $f_n$ is the normal fluid fraction. We expect this fraction to scale with the low-energy quasiparticle density of states, so that in a fully gapped superconductor, $f_n \sim \text{exp}(-\Delta/T)$, while for a clean superconductor with line nodes (such as UPt$_3$), $f_n \sim T/\Delta$ \cite{sigrist-rmp}. In this case, for $T = 100 \; \text{mK}$, we take the normal fluid fraction to be $f_n = 0.1$. 

We take the scattering time to be set by roughly the inverse superconducting gap (appropriate for an unconventional superconductor in the clean limit), $\tau \sim 1/(18 \; \text{GHz}) \sim 60 \;  \text{ps}$. Using these values, the residual normal state dc resistivity in our model can be determined by taking $f_n \goes 1$, yielding $\rho_{\text{dc}} = (\epsi_0 \w_{\text{pl}}^2 \, \tau)^{-1} \approx 2.25  \; \mu \Omega \, \text{cm}$ which is near the actual residual resistivity $\rho_{\text{dc}} \sim 2 \; \mu \Omega \, \text{cm}$ of high quality UPt$_3$ \cite{upt3-rmp,upt3-optics1}.

\bibliography{apssamp}

\end{document}